\documentclass[journal,onecolumn,draftclsnofoot,12pt]{IEEEtran}%
\usepackage{amsfonts}
\usepackage{subfigure}
\usepackage{amsmath}
\usepackage{amssymb}
\usepackage{graphicx}
\usepackage{citesort}%
\newtheorem{theorem}{Theorem}

\newtheorem{definition}{Definition}
\newtheorem{example}{Example}

\newtheorem{lemma}{Lemma}

\newtheorem{proposition}{Proposition}
\newtheorem{remark}{Remark}

\begin{document}

\title{Secrecy capacity of a class of orthogonal relay eavesdropper channels}
\author{Vaneet Aggarwal,~Lalitha Sankar,~A. Robert Calderbank, and H. Vincent Poor
\and \thanks{The authors are with the
Department of Electrical\ Engineering, Princeton University, Princeton, NJ,
USA.

The work of V. Aggarwal and A. R. Calderbank was supported in part by NSF under grant 0701226, by ONR under grant N00173-06-1-G006, and by AFOSR under grant FA9550-05-1-0443. The work of L. Sankar and H. V. Poor was supported in
 part by the National Science Foundation under grant CNS-06-25637.

The material in this paper was presented in part at the Information Theory and Applications Workshop, San Diego, CA, Feb 2009 and at the IEEE International Symposium on Information Theory, Seoul, Korea, Jun 2009.}}
\pubid{~}
\specialpapernotice{~}
\maketitle

\begin{abstract}
The secrecy capacity of relay channels with orthogonal components is studied
in the presence of an additional passive eavesdropper node. The relay and
destination receive signals from the source on two orthogonal channels such that the destination also receives transmissions from the relay on its
channel. The eavesdropper can overhear either one or both of the orthogonal
channels. Inner and outer bounds on the secrecy capacity are developed for
both the discrete memoryless and the Gaussian channel models. For the discrete
memoryless case, the secrecy capacity is shown to be achieved by a
\textit{partial decode-and-forward} (PDF) scheme when the eavesdropper can
overhear only one of the two orthogonal channels. Two new outer bounds are
presented for the Gaussian model using recent capacity results for a Gaussian
multi-antenna point-to-point channel with a multi-antenna eavesdropper. The
outer bounds are shown to be tight for two sub-classes of channels. The first
sub-class is one in which the source and relay are clustered and the
and the eavesdropper receives signals only on the channel from the source and the relay to the destination, for which the PDF
strategy is optimal. The second is a sub-class in which the source does not
transmit to the relay, for which a noise-forwarding strategy is optimal.

\end{abstract}

\section{Introduction}

In wireless networks for which nodes can benefit from cooperation and
packet-forwarding, there is also a need to preserve the confidentiality of
transmitted information from untrusted nodes. Information privacy in wireless networks has
traditionally been the domain of the higher layers of the protocol stack via the use of
cryptographically secure schemes. In his seminal paper on the three-node
wiretap channel, Wyner showed that perfect secrecy of transmitted data from the
source node can be achieved when the physical channel to the eavesdropper is
noisier than the channel to the intended destination, i.e., when
the channel is a degraded broadcast channel \cite{cap_theorems:Wyner_WT}. This
work was later extended by Csisz\'{a}r and K\"{o}rner to all broadcast
channels with confidential messages, in which the source node sends common
information to both the destination and the wiretapper and confidential
information only to the destination \cite{cap_theorems:CK_BCC}. \

Recently, the problem of secure communications has also been studied for a
variety of multi-terminal networks; see, for example,
\cite{cap_theorems:Tekin07,cap_theorems:KhistiWornell,cap_theorems:LiuShamai,cap_theorems:OggHass,cap_theorems:Bloch08,cap_theorems:gracemar,cap_theorems:gracejun,cap_theorems:graceun}, and the references therein. In
\cite{cap_theorems:LaiElGamal}, the authors show that a relay node can
facilitate the transmission of confidential messages from the source to the
destination in the presence of a wiretapper, often referred to as an
eavesdropper in the wireless setting. The authors develop the
rate-equivocation region for this four node relay-eavesdropper channel and
introduce a noise forwarding scheme in which the relay, even if it is unable
to aid the source in its transmissions, transmits codewords independent of the
source to confuse the eavesdropper. A special case where the eavesdropper receives a degraded version of the destination's signal is studied in \cite{cap_theorems:YukselErkip}. In contrast, the relay channel with
confidential messages in which the relay node acts as both a helper and
eavesdropper is studied in \cite{cap_theorems:Oohama06}. Note that in all the three
papers, the relay is assumed to be full-duplex, i.e., it can transmit and
receive simultaneously over the entire bandwidth.

In this paper, we study the secrecy capacity of a relay channel with
orthogonal components in the presence of a passive eavesdropper node. The
orthogonality comes from the fact that the relay and destination receive
signals from the source on orthogonal channels; furthermore, the destination
also receives transmissions from the relay on its (the destination's) channel. The orthogonal
model implicitly imposes a half-duplex transmission and reception constraint
on the relay. For this channel, in the absence of an eavesdropper, El Gamal
and Zahedi showed that a \textit{partial decode-and-forward} (PDF) strategy in
which the source transmits two messages on the two orthogonal channels and the
relay decodes its received signal, achieves the capacity.

We study the secrecy capacity of this channel for both the discrete memoryless
and Gaussian channel models. As a first step towards this, we develop a PDF
strategy for the full-duplex relay eavesdropper channel and extend it to the
orthogonal model. Further, since the eavesdropper can receive signals from
either orthogonal channel or both, three cases arise in the development of the
secrecy capacity. We specialize the outer bounds developed in
\cite{cap_theorems:LaiElGamal} for the orthogonal case and show that for the
discrete memoryless channel, PDF achieves the secrecy capacity for the two
cases where the eavesdropper receives signals in only one of the two
orthogonal channels.

For the Gaussian model, we develop two new outer bounds using recent results
on the secrecy capacity of the Gaussian multiple-input multiple-output
channels in the presence of a multi-antenna eavesdropper (MIMOME) in
\cite{cap_theorems:KhistiWornell,cap_theorems:LiuShamai,cap_theorems:OggHass}.
The first outer bound is a genie-aided bound that allows the source and relay
to cooperate perfectly resulting in a Gaussian MIMOME channel for which
jointly Gaussian inputs maximize the capacity. We show that these bounds are
tight for a sub-class of channels in which the multiaccess channel from the
source and relay to the destination is the bottleneck link, and the eavesdropper is limited to receiving signals on the channel from the source and the relay to the destination. For a
complementary sub-class of channels in which the source-relay link is unusable
due to noise resulting in a \textit{deaf} relay, we develop a genie-aided
bound where the relay and destination act like a two-antenna receiver. We
also show that noise forwarding achieves this bound for this sub-class of channels.

In \cite{cap_theorems:HeYener}, the authors study the secrecy rate of the
channel studied here under the assumption that the relay is co-located with
the eavesdropper and the eavesdropper is completely cognizant of the transmit and
receive signals at the relay. The authors found that using the relay does not
increase the secrecy capacity and hence there is no security advantage to using the
relay. In this paper, we consider the eavesdropper as a separate entity and show
that using the relay increases the secrecy capacity in some cases. In the
model of \cite{cap_theorems:HeYener}, the eavesdropper can overhear only on
the channel to the relay, while we consider three cases in which the
eavesdropper can overhear on either or both the channels.

The paper is organized as follows. In Section \ref{Sec2}, we present the
channel models. In Section \ref{Sec3}, we develop the inner and outer bounds
on the secrecy capacity of the discrete memoryless model. We illustrate these
results with examples in Section \ref{Sec4}. In Section \ref{Sec5}, we present
inner and outer bounds for the Gaussian channel model and illustrate our
results with examples. We conclude in Section \ref{Sec6}.

\section{\label{Sec2}Channel Models and Preliminaries}

\subsection{Discrete Memoryless Model}

A discrete-memoryless relay eavesdropper channel is denoted by $(\mathcal{X}%
_{1}\times\mathcal{X}_{2},p(y,y_{1},y_{2}|x_{1},x_{2}),{\mathcal{Y}}%
\times{\mathcal{Y}}_{1}\times{\mathcal{Y}}_{2})$ such that the inputs to the
channel in a given channel use are $X_{1}\in{\mathcal{X}}_{1}$ and $X_{2}\in{\mathcal{X}}_{2}$ at the
source and relay, respectively, the outputs of the channel are $Y_{1}%
\in{\mathcal{Y}}_{1}$, $Y\in{\mathcal{Y}}$, and $Y_{2}\in{\mathcal{Y}}_{2}$,
at the relay, destination, and eavesdropper, respectively, and the channel
transition probability is given by $p_{YY_{1}Y_{2}|XX_{2}}(y,y_{1}%
,y_{2}|x,x_{2})$ \cite{cap_theorems:LaiElGamal}. The channel is assumed to be
memoryless, i.e. the channel outputs at time $i$ depend only on channel inputs
at time $i$. The source transmits a message $W_{1}\in{\mathcal{W}}%
_{1}=\{1,2,\cdots,M\}$ to the destination using the $(M,n)$ code consisting of

\begin{enumerate}
\item a stochastic encoder $f$ at the source such that $f:{\mathcal{W}}%
_{1}\rightarrow X_{1}^{n}\in{\mathcal{X}}_{1}^{n},$

\item a set of relay encoding functions $f_{r,i}:\left(  Y_{1,1}%
,Y_{1,2},\cdots,Y_{1,i-1}\right)  \rightarrow x_{2,i}$ at every time instant
$i$, and

\item a decoding function at the destination $\Phi:{\mathcal{Y}}%
^{n}\rightarrow{\mathcal{W}}_{1}$.
\end{enumerate}

The average error probability of the code is defined as
\begin{equation}
P_{e}^{n}=\sum_{w_{1}\in{\mathcal{W}}_{1}}\frac{1}{M}\mathbf{Pr}\{\Phi
(Y^{n})\neq w_{1}|w_{1}\text{was sent}\}.
\end{equation}%
\begin{figure}
[ptb]
\begin{center}
\includegraphics[
trim=0.000000in 0.532496in 0.000000in 0.533033in,
height=2.6031in,
width=4.702in
]%
{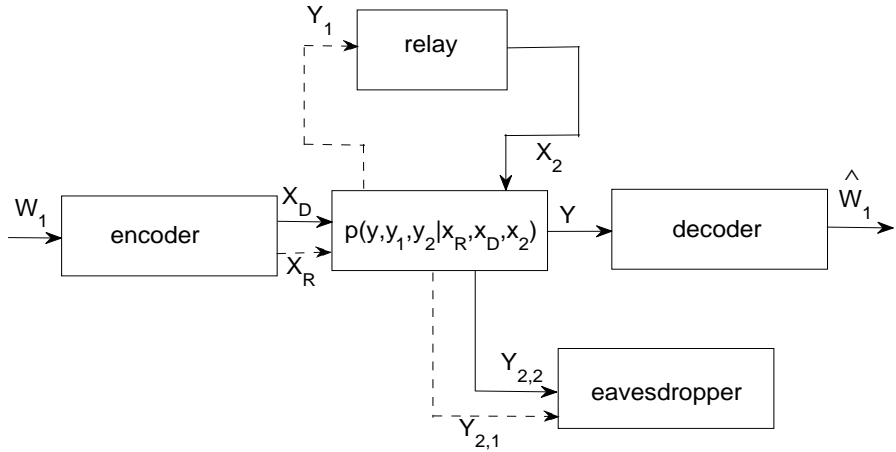}%
\caption{The relay-eavesdropper channel with orthogonal components.}%
\label{Fig_RDCh}%
\end{center}
\end{figure}

The equivocation rate at the eavesdropper is defined as $R_{e}=\frac{1}%
{n}H(W_{1}|{Y}_{2}^{n})$. A perfect secrecy rate of $R_{1}$ is achieved if for any $\epsilon>0$, there exists a sequence of codes $(M,n)$ and an integer $N$ such
that for all $n\geq N$, we have
\begin{align}\label{teststart}
R_{1}  &  =\frac{1}{n}\log_{2}M,\\
P_{e}^{n}  &  \leq\epsilon \text{ and}\\
\frac{1}{n}H(W_{1}|{Y}_{2})  &  \geq R_{1}-\epsilon .\label{testend}
\end{align}
The secrecy capacity is the maximum rate satisfying \eqref{teststart}-\eqref{testend}. The model described above considers a relay that transmits and receives
simultaneously in the same orthogonal channel. Inner and outer bounds for this
model are developed in \cite[Theorem 1]{cap_theorems:LaiElGamal}.

In this paper, we consider a relay eavesdropper channel with orthogonal
components in which the relay receives and transmits on two orthogonal
channels. The source transmits on both channels, one of which is received at
the relay and the other at the destination. The relay transmits along with the
source on the channel received at the destination. Thus, the source signal
$X_{1}$ consists of two parts $X_{R}\in\mathcal{X}_{R}$ and $X_{D}%
\in\mathcal{X}_{D}$, transmitted to the relay and the destination,
respectively, such that ${\mathcal{X}}_{1}={\mathcal{X}}_{D}\times
{\mathcal{X}}_{R}$. The eavesdropper can receive transmissions in one or both
orthogonal channels such that $Y_{2,i}\in{\mathcal{Y}}_{2,i}$ denotes the
received signal at the eavesdropper in orthogonal channel $i$, $i=1,2,$\ and
${\mathcal{Y}}_{2}={\mathcal{Y}}_{2,1}\times{\mathcal{Y}}_{2,2}$. More
formally, the relay eavesdropper channel with orthogonal components is defined as follows.

\begin{definition}
\label{Def_OrthRD}A discrete-memoryless relay eavesdropper channel is said to
have orthogonal components if the sender alphabet ${\mathcal{X}}%
_{1}={\mathcal{X}}_{D}\times{\mathcal{X}}_{R}$ and the channel can be
expressed as
\begin{equation}
p(y,y_{1},y_{2}|x_{1},x_{2})=p(y_{1},y_{2,1}|x_{R},x_{2})\cdot p(y,y_{2,2}%
|x_{D},x_{2}). \label{RED_chpr}%
\end{equation}

\end{definition}

Definition \ref{Def_OrthRD} assumes that the eavesdropper can receive signals
in both channels. In general, the secrecy capacity bounds for this channel
depend on the receiver capabilities of the eavesdropper. To this end,
we explicitly include the following two definitions for the cases in which the
eavesdropper can receive signals in only one of the channels.

\begin{definition}
\label{ChDef1}The eavesdropper is limited to receiving signals on the channel
from the source to the relay, if
\begin{equation}
p(y,y_{1},y_{2,1},y_{2,2}|x_{R},x_{D},x_{2})=p(y_{1},y_{2,1}|x_{R},x_{2})\cdot
p(y|x_{D},x_{2})\cdot p(y_{2,2}).\label{B1_chpr}%
\end{equation}

\end{definition}

\begin{definition}
\label{ChDef2} The eavesdropper is limited to receiving signals on the channel from the
source and the relay to the destination, if
\begin{equation}
p(y,y_{1},y_{2,1},y_{2,2}|x_{R},x_{D},x_{2})=p(y_{1}|x_{R},x_{2})\cdot
p(y,y_{2,2}|x_{D},x_{2})\cdot p(y_{2,1}).\label{B2_chpr}%
\end{equation}

\end{definition}

\begin{remark}
In the absence of an eavesdropper, i.e., for $y_{2,1}=y_{2,2}=0$, the channels
in (\ref{RED_chpr})-(\ref{B2_chpr}) simplify to that of a relay channel with
orthogonal components.
\end{remark}

Thus, depending on the receiver capabilities at the eavesdropper, there are
three cases that arise in developing the secrecy capacity bounds. For brevity,
we henceforth identify the three cases as cases $1$, $2$, and $3$, where cases
1 and 2 correspond to Definitions \ref{ChDef1} and \ref{ChDef2}, respectively,
and case 3 is the general case where the eavesdropper receives signals from
both the channels.

\subsection{Gaussian Model}

For a Gaussian relay eavesdropper channel with orthogonal components, the
signals $Y_{1}$ and $Y$ received at the relay and the destination respectively
in each time symbol $i\in\{1,\cdots,n\}$, are
\begin{equation}
Y_{1}[i]   = h_{s,r}X_{R}[i]+Z_{1}[i]\label{Gsig_rel} \end{equation}
and
\begin{equation}
Y[i]   = h_{s,d}X_{D}[i]+h_{r,d}X_{2}[i]+Z[i]\label{Gsig_dest}%
\end{equation}
where $h_{k,m}$ is the channel gain from transmitter $k\in\{s,r\}$ to receiver
$m\in\{r,d\}$, and where $Z_{1}$ and $Z$ are zero mean unit variance Gaussian random
variables. The transmitted signals $X_{R},$ $X_{D}$, and $X_{2}$ are subject
to average power constraints given by
\begin{equation}%
\begin{array}
[c]{l}%
E[x_{R}^{2}]  \leq P_{R},\\
E[\frac{1}{n}\sum_{i=1}^{n}x_{D}^{2}]  \leq P_{D}, \text{ and }\\
E[\frac{1}{n}\sum_{i=1}^{n}x_{2}^{2}]  \leq P_{2},
\end{array}
\label{GPwr}%
\end{equation}
where $E[.]$ denotes expectation of its argument.
The signals at the eavesdropper are
\begin{align}
Y_{2,1}[i] &  =h_{s,e,1}X_{R}[i]\mathbf{1}_{e,1}+Z_{2,1}[i]\\
Y_{2,2}[i] &  =h_{s,e,2}x_{D}[i]\mathbf{1}_{e,2}+h_{r,e}X_{2}[i]\mathbf{1}%
_{e,2}+Z_{2,2}[i]
\end{align}
where $h_{s,e,1}$ and $h_{s,e,2}$ are the channel gains from the source to
the eavesdropper in the two orthogonal channels, $h_{r,e}$ is the channel gain
from the relay to the eavesdropper, $Z_{2,1}$ and $Z_{2,2}$ are zero-mean unit
variance Gaussian random variables assumed to be independent of the source and relay
signals, and
\[
\mathbf{1}_{e,j}=\left\{
\begin{array}
[c]{ll}%
1 & \text{if the eavesdropper can eavesdrop in orthogonal channel }j=1,2\\
0 & 0\text{ otherwise.}%
\end{array}
\right.
\]
Throughout the sequel, we assume that the channel gains are fixed and known at
all nodes.
%

For a relay channel with orthogonal components, the authors of
\cite{cap_theorems:ElGZahedi01} show that a strategy where the source uses
each channel to send an independent message and the relay decodes the message
transmitted in its channel, achieves capacity. Due to the fact that the relay has
partial access to the source transmissions, this strategy is sometimes also
referred to as \textit{partial decode and forward} (see
\cite{cap_theorems:GKR01}). The achievable scheme involves block Markov
superposition encoding while the converse is developed using the max-flow,
min-cut bounds. The following proposition summarizes this result.

\begin{proposition}
[\cite{cap_theorems:ElGZahedi01}]\label{Prop_ElGZ}The capacity of a relay
channel with orthogonal component is given by
\begin{equation}
C=\max\min\left(  I(X_{R};Y_{1}|X_{2})+I(X_{D};Y|X_{2}),I(X_{R}X_{D}%
X_{2};Y)\right)  \label{Prop1_Cap}%
\end{equation}
where the maximum is over all input distributions of the form
\begin{equation}
p(x_{2})p(x_{R}|x_{2})p(x_{D}|x_{2}). \label{Prop_dist}%
\end{equation}
For the Gaussian model, the bounds in (\ref{Prop1_Cap}) are maximized by
jointly Gaussian inputs transmitting at the maximum power and subject to
(\ref{Prop_dist}).
\end{proposition}

\begin{remark}
While the converse allows for all possible joint distributions of $X_R$, $X_D$, and $X_2$, from the form of the mutual information expressions in (\ref{Prop1_Cap}), it suffices to consider distributions only of the form given by (\ref{Prop_dist}).
%
\end{remark}

We use the standard notation for entropy and mutual information
\cite{cap_theorems:CTbook} and take all logarithms to the base 2 so that our
rate units are bits. For ease of exposition, we write $C\left(  x\right)  $ to
denote $\frac{1}{2}\log\left(  1+x\right) $, and write $x^+$ to denote $\max(x,0)$. We also write random variables with uppercase letters (e.g.
$W_{k}$) and their realizations with the
corresponding lowercase letters (e.g. $w_{k}$). We drop subscripts on probability
distributions if the arguments are lowercase versions of the corresponding random variables.
Finally, for brevity, we henceforth refer to the channel studied here as the
orthogonal relay eavesdropper channel.

\section{\label{Sec3}Discrete memoryless Channel: Outer and Inner Bounds}

In this section, we develop outer and inner bounds for the secrecy
capacity of the discrete-memoryless orthogonal relay eavesdropper channel. The
proof of the outer bounds follows along the same lines as that in
\cite[Theorem 1]{cap_theorems:LaiElGamal} for the full-duplex
relay-eavesdropper channel and is specialized for the orthogonal model
considered here. The following theorem summarizes the bounds for the three
cases in which the eavesdropper can receive in either one or both orthogonal channels.

\begin{theorem}
\label{outerdmc} An outer bound on the secrecy capacity of the relay
eavesdropper channel with orthogonal components is given by%
\begin{equation}%
\begin{array}
[c]{ll}%
\underline{\text{Case}\ 1}: & C_{s}\leq\max\lbrack\min\{I(V_{D}V_{R};YY_{1}%
|V_{2}U),I(V_{D}V_{2};Y|U)\}-I(V_{R};Y_{2}|U)]^{+}\\
\underline{\text{Case}\ 2}: & C_{s}\leq\max\lbrack\min\{I(V_{D}V_{R};YY_{1}%
|V_{2}U),I(V_{D}V_{2};Y|U)\}-I(V_{D}V_{2};Y_{2}|U)]^{+}\\
\underline{\text{Case}\ 3}: & C_{s}\leq\max\lbrack\min\{I(V_{D}V_{R};YY_{1}%
|V_{2}U),I(V_{D}V_{2};Y|U)\}-I(V_{R}V_{D}V_{2};Y_{2}|U)]^{+}%
\end{array}
\label{UBo}%
\end{equation}
where  $U, V_D, V_R$ and $V_2$ are auxiliary random variables, and the maximum is over all joint distributions satisfying $U\rightarrow(V_{R},V_{D},V_{2})\rightarrow(X_{R},X_{D}%
,X_{2})\rightarrow(Y,Y_{1},Y_{2})$.
\end{theorem}

\begin{proof}
The proof is extended from the outer bound in \cite[Theorem 1]%
{cap_theorems:LaiElGamal} to include auxiliary random variables corresponding to
each of the transmitted signals and is developed in Appendix \ref{apouterdmc}.
\end{proof}

Following Proposition \ref{Prop_ElGZ}, a natural question for the
relay-eavesdropper channel with orthogonal components is whether the PDF strategy can achieve the secrecy capacity. To this
end, we first develop the achievable PDF secrecy rates for the class of
\textit{full-duplex} relay-eavesdropper channels and then specialize the
result for the orthogonal model. The following theorem summarizes the inner
bounds on the secrecy capacity achieved by PDF for the full-duplex
(non-orthogonal) relay-eavesdropper channels.

\begin{theorem}
\label{thpdf} An inner bound on the secrecy capacity of a \textit{full-duplex}
relay eavesdropper channel, achieved using partial decode and forward, is
given by
\begin{equation}
C_{s}\geq\min\{I(X_{1};Y|X_{2},V)+I(V;Y_{1}|X_{2}),I(X_{1}X_{2}V;Y)\}-I(X_{1}%
X_{2};Y_{2})
\end{equation}
for all joint distributions of the form
\begin{equation}
p(v)p(x_{1}|v)p(x_{2}|v)p(y_{1},y|x_{1},x_{2}).
\end{equation}

\end{theorem}

\begin{proof}
The proof is developed in Appendix \ref{appdg} and uses block Markov
superposition encoding at the source such that in each block, the relay
decodes a part of the source message while the eavesdropper has access to both
source messages.
\end{proof}

The following theorem specializes Theorem \ref{thpdf} for the orthogonal
relay-eavesdropper channel.

\begin{theorem}
\label{thach} An inner bound on the secrecy capacity of the orthogonal relay
eavesdropper channel, achieved using partial decode and forward over all joint distributions of the form $p(x_{R},x_{D},x_{2})$, is given by%
\begin{equation}%
\begin{array}
[c]{ll}%
\underline{\text{Case}\ 1}: & C_{s}\geq\min\{I(X_{D}X_{R};YY_{1}%
|X_{2}),I(X_{D}X_{2};Y)\}-I(X_{R};Y_{2})\\
\underline{\text{Case}\ 2}: & C_{s}\geq\min\{I(X_{D}X_{R};YY_{1}%
|X_{2}),I(X_{D}X_{2};Y)\}-I(X_{D},X_{2};Y_{2})\\
\underline{\text{Case}\ 3}: & C_{s}\geq\min\{I(X_{D}X_{R};YY_{1}%
|X_{2}),I(X_{D}X_{2};Y)\}-I(X_{R};Y_{2}|X_{2})-I(X_{D},X_{2};Y_{2})
\end{array}
\label{IBo}%
\end{equation}

\end{theorem}

\begin{proof}
The proof is developed in Appendix \ref{apach} and involves specializing the
bounds in Theorem \ref{thpdf} for the orthogonal model. It is further shown
that the input distribution can be generalized to all joint probability
distributions $p(x_{R},x_{D},x_{2})$.
\end{proof}

The bounds in (\ref{IBo}) can be generalized by randomizing the channel
inputs. We now prove that PDF with randomization achieves the secrecy capacity.

\begin{theorem}
\label{thaccap} The secrecy capacity of the relay channel with orthogonal
complements is
\begin{equation}%
\begin{array}
[c]{ll}%
\underline{\text{Case}\ 1}: & C_{s}=\max[\min\{I(V_{D}V_{R};YY_{1}|V_{2}%
U),I(V_{D}V_{2};Y|U)\}-I(V_{R};Y_{2}|U)]^{+}\\
\underline{\text{Case}\ 2}: & C_{s}=\max[\min\{I(V_{D}V_{R};YY_{1}|V_{2}%
U),I(V_{D}V_{2};Y|U)\}-I(V_{D}V_{2};Y_{2}|U)]^{+}\\
\underline{\text{Case}\ 3}: & C_{s}\leq\max\lbrack\min\{I(V_{D}V_{R};YY_{1}%
|V_{2}U),I(V_{D}V_{2};Y|U)\}-I(V_{R}V_{D}V_{2};Y_{2}|U)]^{+}%
\end{array}
\end{equation}
where  $U, V_D, V_R$ and $V_2$ are auxiliary random variables, and the maximum is over all joint distributions satisfying $U\rightarrow(V_{R},V_{D},V_{2})\rightarrow(X_{R},X_{D}%
,X_{2})\rightarrow(Y,Y_{1},Y_{2})$. Furthermore, for Case 3,
\begin{equation}
C_{s}\geq\lbrack\min\{I(V_{D}V_{R};YY_{1}|V_{2}U),I(V_{D}V_{2};Y|U)\}-I(V_{R}%
;Y_{2}|V_{2}U)-I(V_{D},V_{2};Y_{2}|U)]^{+}%
\end{equation}
for all joint distributions satisfying $U\rightarrow(V_{R},V_{D},V_{2})\rightarrow(X_{R},X_{D},X_{2}%
)\rightarrow(Y,Y_{1},Y_{2})$.
\end{theorem}

\begin{proof}
The upper bounds follow from Theorem \ref{outerdmc}. For the lower bound, we
prefix a memoryless channel with inputs $V_{R},$ $V_{D},$ and $V_{2}$ and
transition probability $p(x_{R},x_{D},x_{2}|v_{R},v_{D},v_{2})$ (this prefix can potentially increase the achievable secrecy rates as in \cite{cap_theorems:CK_BCC,cap_theorems:LaiElGamal}). The
time-sharing random variable $U$ ensures that the set of achievable rates is convex.
\end{proof}

\begin{remark}
In contrast to the non secrecy case, where the orthogonal channel model
simplifies the cut-set bounds to match the inner PDF bounds, for the
orthogonal relay-eavesdropper model in which the eavesdropper receives in both
channels, i.e., when the orthogonal receiver restrictions at the relay and
intended destination do not apply to the eavesdropper, in general, the outer
bound can be strictly larger than the inner PDF bound.
\end{remark}

In the following section, we illustrate these results with three examples.

\section{\label{Sec4}Examples}

\begin{example}
\label{Eg1}Consider a orthogonal relay eavesdropper channel with
$\mathcal{X}_{R}=\mathcal{X}_{D}=\mathcal{X}_{2}=\left\{  0,1\right\}  $. The
outputs at the relay and destination are given by
\begin{align}
Y_{1}  &  =X_{R} \text{\ \ and}\\
Y  &  =X_{D}X_{2},
\end{align}
while the output at the eavesdropper is
\begin{equation}%
\begin{array}
[c]{ll}%
Y_{2,1}=X_{R} & (\text{channel }1) \text{\ \ and}\\
Y_{2,2}=\left\{
\begin{array}
[c]{l}%
1\text{ if }X_{D}\leq X_{2}\\
0\text{ otherwise }%
\end{array}
\right.  & (\text{channel }2).
\end{array}
\end{equation}
Since the destination can receive at most 1 bit in every use of the channel,
the secrecy capacity of this channel is at most $1$ bit per channel use. We
now show that this secrecy capacity can be achieved. In each channel use, let
the source send bit $w\in\{0,1\}$ such that $X_{R}=0$, $X_{D}=w$, and
$X_{2}=1$. Since $X_{2}=1$, the receiver obtains $w$ while the eavesdropper
receives $Y_{2,1}=0$ and $Y_{2,2}=1$ irrespective of the value of bit $w$.
Hence, a perfect secrecy capacity of $1$ can be achieved.
\end{example}

The code design in Example \ref{Eg1} did not require randomization. We now
present an example where randomization is necessary.

\begin{figure}[ptbh]
\centering
\subfigure[Example \ref{Ex2}]{
\includegraphics[
trim=0.000000in 0.000000in 0.939081in 0.000000in,
height=2.2in,
]{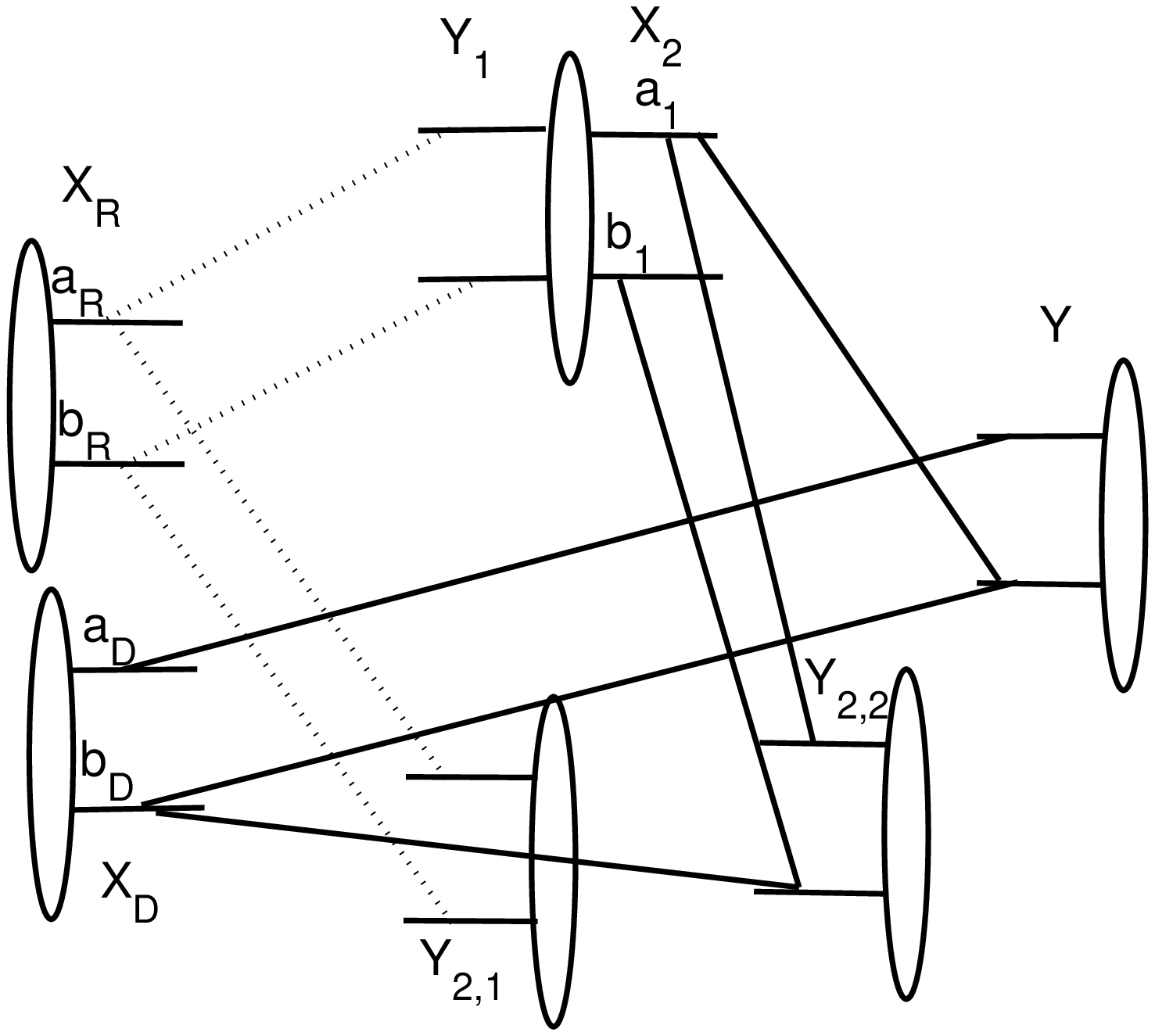}\label{Fig_Ex2}}\hspace{.5cm} \subfigure[Example \ref{Ex3}]{
\includegraphics[
trim=0.000000in 0.000000in 0.805904in 0.000000in,
height=2.2in,
]{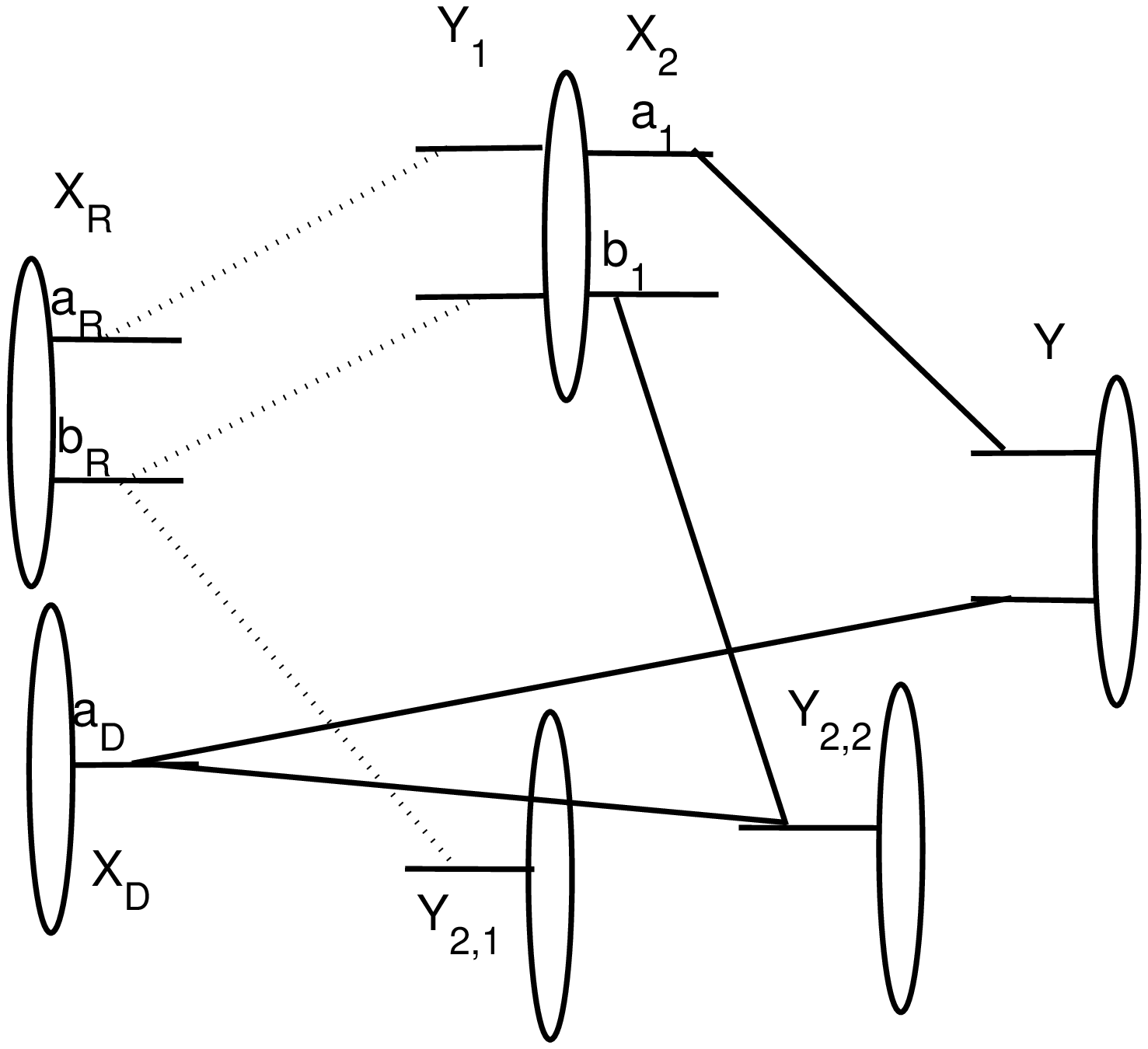}\label{Fig_Ex3}}\caption{Orthogonal relay eavesdropper channel model
of Examples \ref{Ex2} and \ref{Ex3}.}%
\end{figure}

\begin{example}
\label{Ex2}Consider an orthogonal relay eavesdropper channel where all the
input and output alphabets are the same and given by $\{0,1\}^{2}$. We write
$X_{R}=(a_{R},b_{R})$, $X_{D}=(a_{D},b_{D})$, and $X_{2}=(a_{1},b_{1})$ to
denote the vector binary signals at the source and the relay. The outputs of
this channel, shown in Figure \ref{Fig_Ex2}, at the relay, destination, and
the eavesdropper are given by%
\begin{align}
Y  &  =(a_{D},b_{D}\oplus a_{1}),\\
Y_{1}  &  =(a_{R},b_{R}),\\
Y_{2,1}  &  =(a_{R},b_{R}) \text{\ \ and}\\
Y_{2,2}  &  =(a_{1},b_{1}\oplus a_{D}),
\end{align}
where $\oplus$ denotes the binary XOR operation. The capacity of this channel
is at most $2$ bits per channel use as the destination, via $Y$, can receive
at most $2$ bits per channel use. We will now show that a secrecy capacity of
$2$ bits per channel use can be achieved. Consider the following coding
scheme. In every channel use, the relay flips an unbiased coin to generate a
bit $n\in\{0,1\}$ such that its transmitted signal is%
\begin{equation}
X_{2}=(0,n).\nonumber
\end{equation}
In every use of the channel, the source transmits $2$ bits, denoted as $w_{1}$
and $w_{2}$, using%
\begin{align}
X_{R}  &  =(0,0) \text{\ \ and}\nonumber\\
X_{D}  &  =(w_{1},w_{2}).\nonumber
\end{align}
For these transmitted signals, the receiver and eavesdropper receive
\begin{align}
Y  &  =(w_{1},w_{2}),\\
Y_{2,1}  &  =(0,0)\text{ \ \ and}\\
Y_{2,2}  &  =(0,n\oplus w_{1}).
\end{align}
Thus, the receiver receives both bits while the eavesdropper is unable to
decode any information due to the randomness of $n$. This is an example where
transmitting a random code from the relay is required to achieve the secrecy capacity.
\end{example}

In the above two examples, the source to relay link was completely available
to the eavesdropper and hence the relay could at best be just used to send
random bits. In the next example, we show that the secrecy capacity is
achieved by the relay transmitting a part of the message as well as a random signal.

\begin{example}
\label{Ex3}Consider an orthogonal relay eavesdropper channel where the input
and output signals at the source, relay, and destination are binary two-tuples
while $\mathcal{Y}_{2,1}$ and $\mathcal{Y}_{2,2}$ at the eavesdropper are
binary alphabets. We write $X_{R}=(a_{R},b_{R})$, $X_{D}=a_{D}$ and
$X_{2}=(a_{1},b_{1})$ to denote the vector binary signals at the source and
the relay. The outputs at the relay, destination and the eavesdropper are also
vector binary signals given by
\begin{align}
Y  &  =(a_{1},a_{D}),\\
Y_{1}  &  =(a_{R},b_{R}),\\
Y_{2,1}  &  =(b_{R}) \text{\ \ and}\\
Y_{2,2}  &  =(b_{1}\oplus a_{D}),
\end{align}
\newline as shown in Figure \ref{Fig_Ex3}. As in the previous example, the
capacity of this channel is also at most $2$ bits per channel use. We now show
that a secrecy capacity of $2$ bits per channel use can be achieved for this
example channel. Consider the following coding scheme: in the $i^{th}$ use of
the channel, the source encodes $2$ bits, denoted as $w_{1,i}$ and $w_{2,i}$
as
\begin{align}
X_{R}  &  =(w_{1,i},0) \text{\ \ and}\nonumber\\
X_{D}  &  =(w_{2,i}).\nonumber
\end{align}
The relay receives $w_{1,i-1}$ in the previous use of the channel.
Furthermore, in each channel use, it also generates a uniformly random bit
$n_{i}$, and transmits
\begin{equation}
X_{2}=(w_{1,i-1},n_{i}).
\end{equation}
With these transmitted signals, the received signals at the receiver and the
eavesdropper are%
\begin{align}
Y  &  =(w_{1,i-1},w_{2,i}),\\
Y_{2,1}  &  =(0) \text{\ \ and}\\
Y_{2,2}  &  =(n_{i}\oplus w_{2,i}).
\end{align}
Thus, over $n+1$ uses of the channel the destination receives all $2n+1$ bits
transmitted by the source. On the other hand, in every use of the channel, the
eavesdropper cannot decode either source bit.
\end{example}

\section{\label{Sec5}Gaussian Model}

\subsection{Inner and Outer Bounds}

We now develop inner and outer bounds for the Gaussian orthogonal relay
eavesdropper channel. Determining the optimal input distribution for all the
auxiliary random variables in the outer bounds in Theorem \ref{thaccap} is not
straightforward. To this end, we develop new outer bounds using a recent
result on the secrecy capacity of the class of Gaussian multiple input,
multiple output, multi-antenna eavesdropper channels (see
\cite{cap_theorems:KhistiWornell,cap_theorems:LiuShamai,cap_theorems:OggHass}%
). The class of MIMOME channels is characterized by a single source with an
$m\times1$ vector input $\mathbf{X}$ and $k\times1$ and $t\times1$ vector
outputs $\mathbf{Y}$ and $\mathbf{Y}_{e}$ at the intended destination and
eavesdropper, respectively, given by%
\begin{equation}%
\begin{array}
[c]{c}%
\mathbf{Y}[i]=\mathbf{HX}[i]+\mathbf{Z}[i] \text{\ \ and}\\
\mathbf{Y}_{e}[i]=\mathbf{H}_{e}\mathbf{X}[i]+\mathbf{Z}_{e}[i]
\end{array}
\label{MME_ch}%
\end{equation}
where in every channel use $i$, $\mathbf{Z}[i]$ and $\mathbf{Z}_{e}[i]$ are
zero-mean Gaussian vectors with identity covariance matrices that are
independent across time symbols. The channel input satisfies an average
transmit power constraint:
\begin{equation}
\frac{1}{n}\sum\nolimits_{i=1}^{n}\left\Vert \mathbf{x}\right\Vert ^{2}\leq
P\mathbf{.} \label{MME_Pwr}%
\end{equation}
In applying the multi-antenna secrecy capacity results, we develop an outer
bound in which the source and relay are modeled jointly as a multi-antenna
transmitter. However, unlike the average power constraint for the
MIMOME\ channels in (\ref{MME_Pwr}), our outer bound requires a per antenna
power constraint. To this end, we apply the results developed in
\cite{cap_theorems:LiuShamai} in which a more general transmitter covariance
constraint is considered such that
\begin{equation}
\frac{1}{n}\sum\nolimits_{i=1}^{n}\left(  \mathbf{x}\left[  i\right]
\mathbf{x}^{T}\left[  i\right]  \right)  \preceq\mathbf{S} \label{MME_Cov}%
\end{equation}
where $\mathbf{S}$ is a positive semidefinite matrix and $A\preceq B$ denotes that $B-A$ is a positive semidefinite matrix. The secrecy capacity of
this channel is summarized in the following theorem.


\begin{lemma}
[\cite{cap_theorems:LiuShamai}]\label{le:mimome} The secrecy capacity of the
MIMOME channel of (\ref{MME_ch}) subject to (\ref{MME_Cov}) is given by
\begin{equation}
C_{s}=\max_{0\preceq\mathbf{K}_{\mathbf{X}}\preceq\mathbf{S}}\left(  \frac
{1}{2}\log\det\left(  {\mathsf I}+\mathbf{H\mathbf{K}_{X}H}^{T}\right)  -\frac{1}{2}%
\log\det\left(  {\mathsf I}+\mathbf{H}_{e}\mathbf{K}_{\mathbf{X}}\mathbf{H}_{e}%
^{T}\right)  \right)  . \label{MME_Cap}%
\end{equation}

\end{lemma}

\begin{remark}
The expression in (\ref{MME_Cap}) can also be written as
\begin{equation}
C_{s}=\max[I(\mathbf{X}^{\ast}\mathbf{;Y})-I(\mathbf{X}^{\ast};\mathbf{Y}_{e})]
\label{MME_Cap_alt}%
\end{equation}
where he maximum is over all $\mathbf{X}^{\ast}\sim\mathcal{N}(0,\mathbf{K}_{\mathbf{X}}).$
\end{remark}

We now present an outer bound on the Gaussian orthogonal relay eavesdropper
channel using Lemma \ref{le:mimome}.

\begin{theorem}
\label{thgaussub} An outer bound on the secrecy capacity of the Gaussian
orthogonal relay eavesdropper channel is given by
\begin{subequations}\label{GOB1}%
\begin{eqnarray}%
\underline{\text{Case}\ 1} : & C_{s}  \leq & \max[I(X_{D}X_{2};Y)-I(X_{R};Y_{2})]\label{GOB1a}\\
\underline{\text{Case}\ 2} : & C_{s}  \leq & \max[I(X_{D}X_{2};Y)-I(X_{D}X_{2};Y_{2})]\label{GOB1b}\\
\underline{\text{Case}\ 3} : & C_{s}  \leq & \max[I(X_{D}X_{2};Y)-I(X_{R}X_{D}X_{2};Y_{2})]\label{GOB1c}
\end{eqnarray}
\end{subequations}
where the maximum is over all $[X_{R}$ $X_{D}$ $X_{2}]^{T}\sim\mathcal{N}(0,\mathbf{K}_{\mathbf{X}})$
where $\mathbf{K}_{\mathbf{X}}=E[\mathbf{XX}^{T}]$ has diagonal entries that
satisfy (\ref{GPwr}).
\end{theorem}
\begin{remark}
In \eqref{GOB1a} and \eqref{GOB1c}, the $X_R^*$ maximizing the outer bound on the secrecy capacity is $X_R^* =0$. On the other hand, $X_R^*$ can be chosen to be arbitrary for \eqref{GOB1b}.

\end{remark}
\begin{proof}
An outer bound on the secrecy capacity of the relay eavesdropper channel
results from assuming that the source and relay can cooperate over a noiseless
link without causality constraints. Under this assumption, the problem reduces
to that of a MIMOME channel. Thus, applying Lemma \ref{le:mimome} and using
the form in (\ref{MME_Cap_alt}), for $\mathbf{X}=[X_{R}$ $X_{D}$ $X_{2}%
]^{T}\sim\mathcal{N}(0,\mathbf{K}_{\mathbf{X}})$, the secrecy capacity can be
upper bounded as%
\begin{align}
C_{s}  &  \leq \max[I(X_{R}X_{D}X_{2};Y)-I(X_{R}X_{D}X_{2};Y_{2})]\\
&  =\max[I(X_{D}X_{2};Y)+I(X_{R};Y|X_{D}X_{2})-I(X_{R}X_{D}X_{2};Y_{2})]\\
&  =\max[I(X_{D}X_{2};Y)-I(X_{R}X_{D}X_{2};Y_{2})] \label{GOB_pr}%
\end{align}
where (\ref{GOB_pr}) follows from the orthogonal model in (\ref{RED_chpr}).
Finally, applying the conditions on the eavesdropper receiver for the three
cases simplifies the bounds in (\ref{GOB_pr}) to (\ref{GOB1}).
\end{proof}

The PDF inner bounds developed in Section III for the discrete memoryless case can be applied to
the Gaussian model with Gaussian inputs at the source and relay. In fact, for
all three cases, the inner bounds require taking a minimum of two rates, one
achieved jointly by the source and relay at the destination and the other
achieved by the source at the relay and destination. Comparing the inner
bounds in (\ref{IBo}) with the outer bounds in (\ref{GOB1}), for those
channels in which the source and relay are clustered close enough that the
bottle-neck link is the combined source-relay link to the destination and the
eavesdropper overhears only the channel from the source and the relay to the destination,  the
secrecy capacity can be achieved. This is summarized in the following theorem.

\begin{theorem}
For a class of \textit{clustered} orthogonal Gaussian relay channels with%
\begin{equation}
I(X_{D}X_{2};Y)<\max_{p(x_R|x_D,x_2)}I(X_{D}X_{R};YY_{1}|X_{2}),
\end{equation}
 the secrecy capacity for case 2 is achieved by PDF and
is given by%
\begin{equation}
\begin{array}
[c]{ll}%
\underline{\text{Case}\ 2}: & C_{s}=\max[I(X_{D}X_{2};Y)-I(X_{D},X_{2};Y_{2})]
\end{array}
\end{equation}
where the maximum is over $\mathbf{X}=[X_{R}$ $X_{D}$ $X_{2}]^T\sim\mathcal{N}(0,\mathbf{K}%
_{\mathbf{X}})$.
\end{theorem}

For a relay channel without secrecy constraints, the cut-set outer bounds are
equivalent to two multiple-input multiple-output (MIMO) bounds, one that results from assuming a noiseless
source-relay link and the other that results from assuming a noiseless
relay-destination link. Under a secrecy constraint, the outer bound in Theorem
\ref{thgaussub} is based on the assumption of a noiseless source-relay link.
The corresponding bound with a noiseless relay-destination link remains unknown.

We now consider a sub-class of Gaussian orthogonal relay eavesdropper channels
for which $h_{s,r}=0.$ For this sub-class, the source does not send any
messages on channel $1$, i.e., $X_{R}=0$. Such a sub-class is a subset of a
larger sub-class of channels with very noisy unreliable links from the source
to the relay. We present an upper bound on the secrecy capacity for this
sub-class and show that the noise-forwarding strategy introduced in
\cite{cap_theorems:LaiElGamal} achieves this outer bound. Central to our
proof is an additional constraint introduced in developing the outer bounds on the eavesdropper that it does not decode
the relay transmissions. Clearly, limiting the eavesdropper capabilities can
only improve the secrecy rates, and thus, an outer bound for this channel with
a constrained eavesdropper is also an outer bound for the original channel
(with $h_{s,r}=0$ in both cases) with an unconstrained eavesdropper. We show
that the outer bound for the constrained channel can be achieved by the
strategy of noise-forwarding developed for the unconstrained channel.

\begin{theorem}
\label{thgaussubb10} The secrecy capacity of a sub-class of Gaussian
orthogonal relay eavesdropper channels with $h_{s,r}=0$ for Cases $2$ and
$3$ is given by
\begin{multline}
C_{s}=\max_{E[X_{D}]^{2}\le P_D, E[X_{2}]^{2}\le P_2}\min\left\{C\left(  |h_{s,d}|^{2}E[X_{D}^{2}]+|h_{r,d}|^{2}E[X_{2}%
^{2}]\right)  -C\left(  |h_{s,e,2}|^{2}E[X_{D}^{2}]+|h_{r,e}|^{2}E[X_{2}%
^{2}]\right)  ,\right.\\
\text{ \ \ \ \ \ \ \ \ \ \ \ }\left.C(|h_{s,d}|^{2}E[X_{D}^{2}])-C\left(
|h_{s,e,1}|^{2}E[X_{D}^{2}]/( 1+|h_{r,e}|^{2}E[X_{2}^{2}]) \right)
\right\}.%
\end{multline}
\end{theorem}

\begin{proof}
\noindent\textit{Outer Bound}: An outer bound on the secrecy capacity is
obtained by applying Theorem \ref{thgaussub} for Cases 2 and 3 as
\begin{align}
C_{s}  &  \leq \max[I(X_{D}X_{2};Y)-I(X_{D}X_{2};Y_{2})]\label{NFOB_1}\\
&  =\max_{E[X_{D}]^{2}\le P_D, E[X_{2}]^{2}\le P_2} [C(|h_{s,d}|^{2}E[X_{D}^{2}]+|h_{r,d}|^{2}E[X_{2}^{2}])-C(|h_{s,e,2}%
|^{2}E[X_{D}^{2}]+|h_{r,e}|^{2}E[X_{2}^{2}])] \label{NFOB_2}%
\end{align}
where (\ref{NFOB_2}) holds because  $h_{s,r} = 0$ implies $X_R = 0$.  This follows from the fact that due to a lack of a communication link between the source and the relay, i.e., $h_{s,r} = 0$, the relay is oblivious to the source transmissions. Since the relay and the source do not share common randomness, one can set $X_R = 0$. Further, since $X_2$ depends on $X_D$ only via $X_R$ and $X_R=0$, $X_2$ is independent of $X_D$. Finally, the optimality of Gaussian signaling follows from Theorem \ref{thgaussub}.

We now develop a second outer bound under the assumption that the relay and
the destination have a noiseless channel such that they act like a two-antenna
receiver. One can alternately view this as an improved channel that results
from having a genie that shares perfectly the transmitted and received signals
at the relay with the destination. Since $X_{2}$ is independent of $X_{D}$,
the destination can perfectly cancel $X_{2}$ from its received signal, and
thus, from (\ref{Gsig_dest}), the effective received signal at the destination
can be written as
\begin{equation}
Y^{\prime}=h_{s,d}X_{D}+Z.
\end{equation}
On the other hand for the constrained eavesdropper, since the relay's signal $X_2$ acts as interference and is independent of $X_D$,
the information received at the eavesdropper is minimized when $X_2$ is the worst case noise, i.e., when it is Gaussian distributed \cite[Theorem II.1]{cap_theorems:Diggavi01}. The equivalent signal received at the eavesdropper
is then
\begin{equation}
Y_{2,2}^{\prime}=h_{s,e,2}X_{D}+\sqrt{|h_{r,e}|^{2}E[X_{2}^{2}]+1}%
Z_{2,2}^{\prime}%
\end{equation}
where $Z_{2,2}^{\prime}$ is Gaussian with zero mean and unit variance. Thus,
the constrained eavesdropper channel simplifies to a MIMOME channel with a
single-antenna source transmitting $X_{D}$ and single-antenna receiver and
eavesdropper receiving $Y^{\prime}$ and $Y_{2,2}^{\prime}$, respectively. For
this channel, from Lemma \ref{le:mimome}, the secrecy capacity of this
constrained eavesdropper channel is upper bounded as%
\begin{equation}
C_{s}\leq \max_{E[X_{D}]^{2}\le P_D, E[X_{2}]^{2}\le P_2}[C(|h_{s,d}|^{2}E[X_{D}^{2}])-C\left( |h_{s,e,1}%
|^{2}E[X_{D}^{2}] /( 1+|h_{r,e}|^{2}E[X_{2}^{2}]) \right)] . \label{NFOB}%
\end{equation}
Finally, since (\ref{NFOB}) is an upper bound for the channel with an
eavesdropper constrained to ignore $X_{2}$, it is also an upper bound for the
channel in which the eavesdropper is not constrained.

\textit{Inner Bound}: The lower bound follows from the noise forwarding
strategy introduced in \cite[Theorem 3]{cap_theorems:LaiElGamal}. In this strategy, the relay sends
codewords independent of the source message, which helps in confusing the
eavesdropper. The noise forwarding strategy transforms the relay-eavesdropper
channel into a compound multiple access channel, where the source/relay to the
receiver is the first multiple access channel and the source/relay to the
eavesdropper is the second one.
\end{proof}

\subsection{Illustration of Results}

\begin{figure}[ptbh]
\label{eve15} \centering
\subfigure[Case 1.]{
\includegraphics[height=2in]{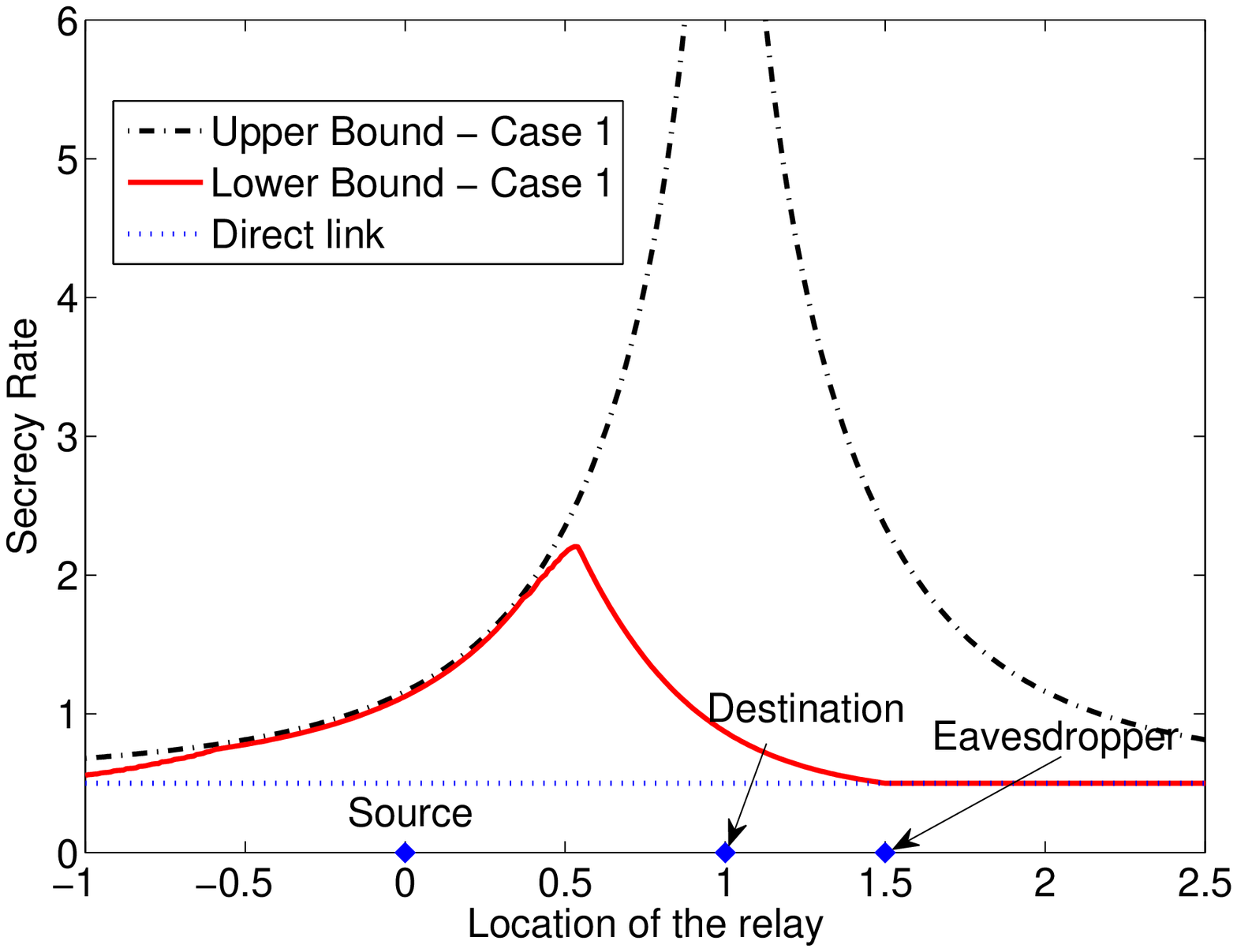}
\label{eve15b1}
} \subfigure[Case 2.]{
\includegraphics[height=2in]{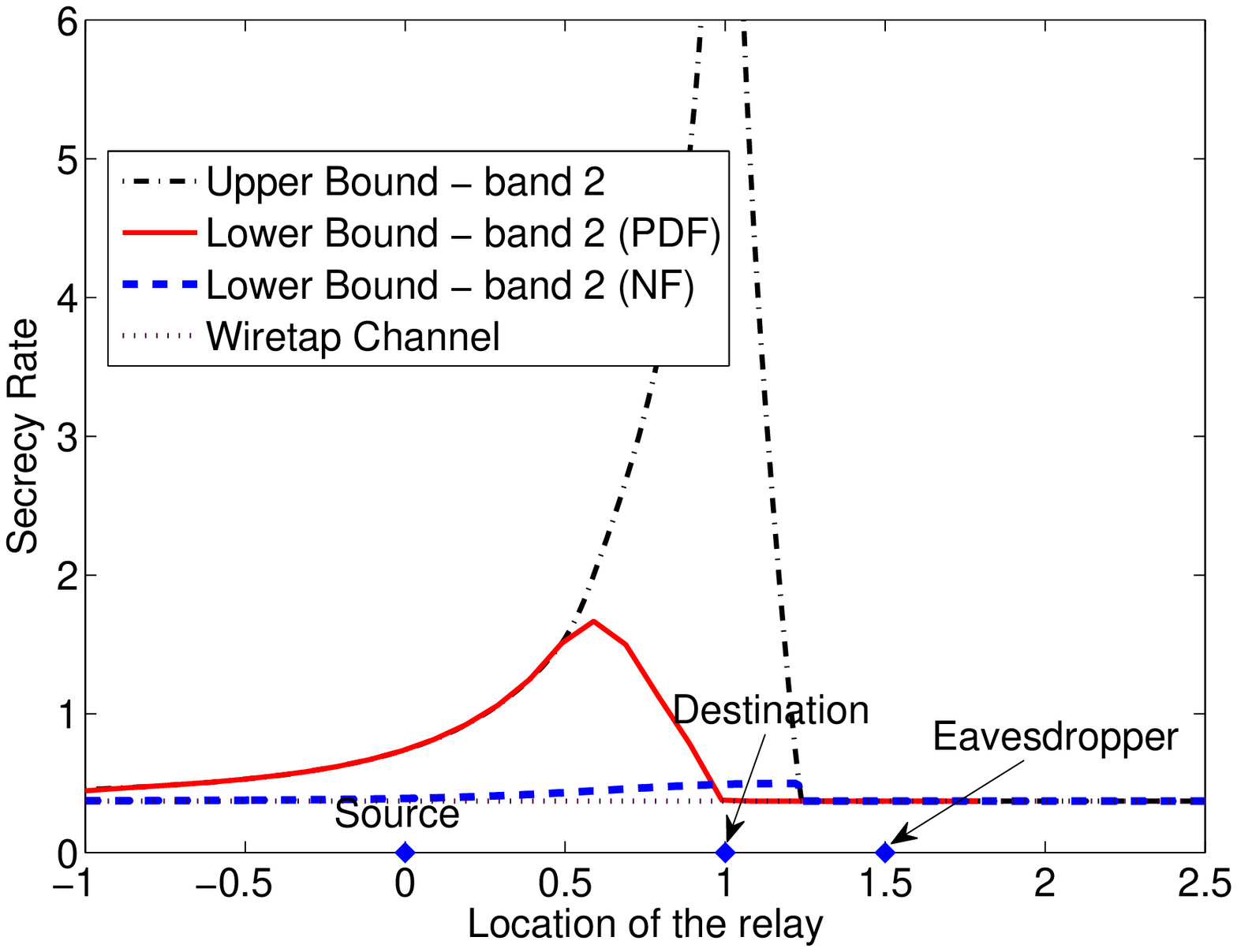}
\label{eve15b2}
} \subfigure[Case 3.]{
\includegraphics[height=2in]{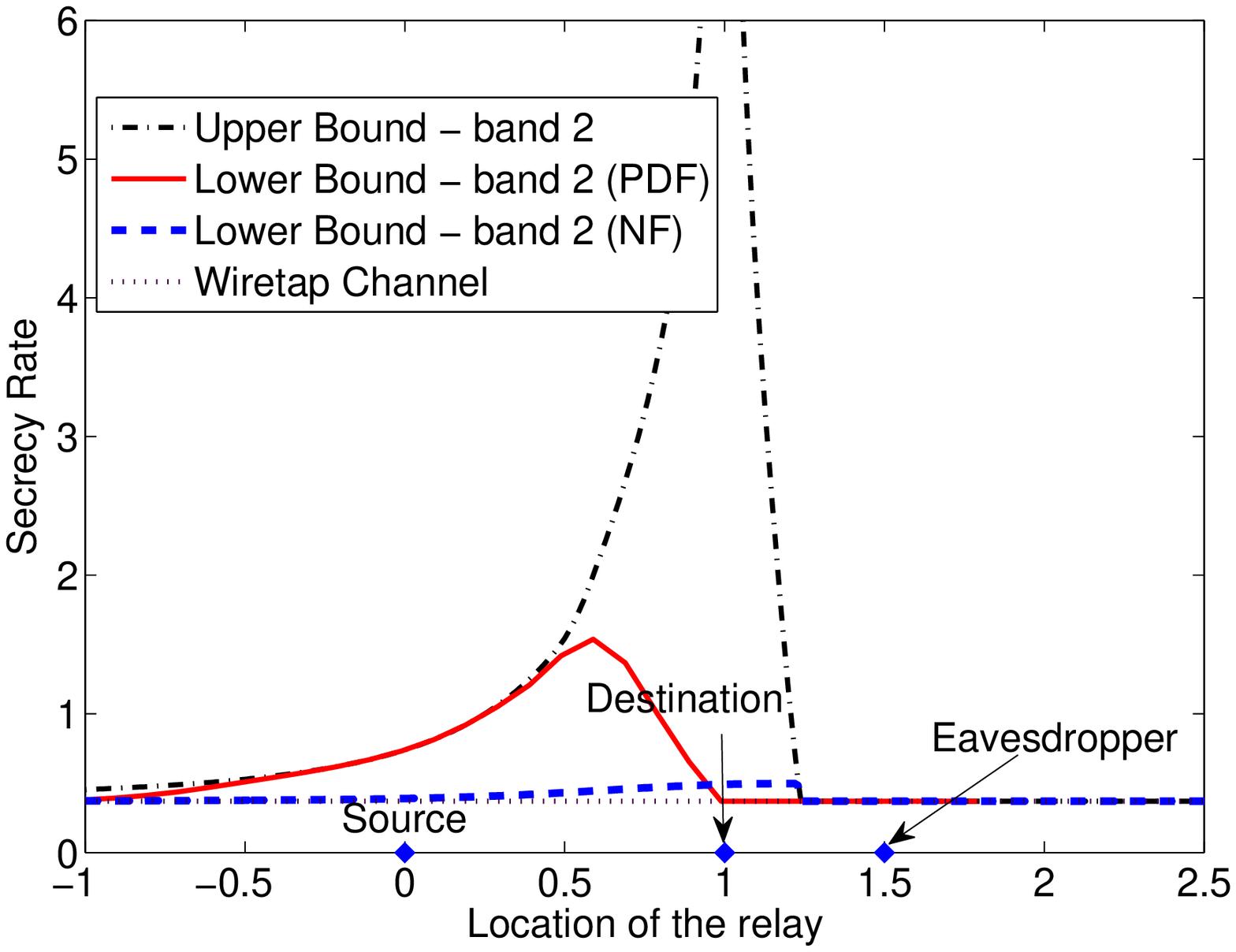}
\label{eve15b12}
}\caption{Source is at $\left(0,0\right)$, destination at $\left(1,0\right)$ and eavesdropper is at
$\left(1.5,0\right)$. A distance fading model with $\alpha=2$ is taken and power constraints for
$X_{R}$, $X_{D}$ and $X_{2}$ are all unity.}%
\end{figure}

We illustrate our results for the Gaussian model for a class of linear
networks in which the source is placed at the origin and the destination is unit
distance from the source at $\left(  1,0\right)  $. The eavesdropper is
at $\left(1.5,0\right)$. The channel gain $h_{m,k}$, between transmitter $m$ and receiver
$k$, for each $m$ and $k$, is modeled as a distance dependent path-loss gain
given by
\begin{equation}%
\begin{array}
[c]{cc}%
h_{m,k}=\frac{1}{d_{m,k}^{\alpha/2}} & \text{for all }m\in\left\{
s,r\right\}  ,k\in\left\{  r,d,e\right\}
\end{array}
\end{equation}
where $\alpha$ is the path-loss exponent. The maximum achievable PDF secrecy
rate is plotted as a function of the relay position along the line connecting
the source and the eavesdropper as shown in Figure \ref{eve15}. Furthermore, as a baseline assuming the relay does not transmit, i.e., $X_R = 0$, the secrecy capacity of the resulting direct link and the wire-tap channel for cases $2$ and $3$, respectively, are included in all three plots in Fig. \ref{eve15}. The rates are
plotted in separate sub-figures for the three cases in which the eavesdropper
receives signals in only one or both channels. In all cases, the path loss
exponent $\alpha$ is set to $2$ and the average power constraint on $X_{R},$
$X_{D}\,$, and $X_{2}$ is set to unity. In addition to PDF, the secrecy rate
achieved by noise forwarding (NF) is also plotted.

In Fig \ref{eve15}, for all three cases, the PDF secrecy rates are obtained by choosing
the input signal $\mathbf{X}=[X_{R}$ $X_{D}$ $X_{2}]^{T}$ to be Gaussian
distributed and numerically optimizing the rates over the covariance matrix $\mathbf{K}%
_{\mathbf{X}}=E[\mathbf{X}\mathbf{X}^{T}]$ (more precisely the three variances of $X_R$, $X_D$, $X_2$ and the pairwise correlation among these three variables). We observe that the numerical results match the theoretical capacity result for Case 2 that PDF is optimal when the relay is close to the source. Further, the upper bounds for Case 2 and  Case 3 are the same as seen also in \eqref{GOB1b}-\eqref{GOB1c}. On the other hand, when the
relay is farther away than the eavesdropper and destination are from the
source, there are no gains achieved by using the relay relative to the
non-relay wiretap secrecy capacity. Finally, for cases $2$ and $3$, NF
performs better than PDF when the relay is closer to the destination.

\section{\label{Sec6}Conclusions}

We have developed bounds on the secrecy capacity of relay eavesdropper
channels with orthogonal components in the presence of an additional passive
eavesdropper for both the discrete memoryless and Gaussian channel models. Our
results depend on the capability of the eavesdropper to overhear either or
both of the two orthogonal channels that the source uses for its
transmissions. For the discrete memoryless model, when the eavesdropper is
restricted to receiving in only one of the two channels, we have shown that
the secrecy capacity is achieved by a partial decode-and-forward strategy.

For the Gaussian model, we have developed a new outer bound using recent
results on the secrecy capacity of Gaussian MIMOME channels. When the eavesdropper is restricted to overhearing on the channel from the source and the relay to the destination, our bound is tight for a sub-class of channels where the source and the relay are clustered such that the combined link from the source and the relay to the destination is the bottleneck. Furthermore, for a sub-class where the
source-relay link is not used, we have developed a new MIMOME-based outer
bound that matches the secrecy rate achieved by the noise forwarding strategy.

A natural extension to this model is to study the secrecy capacity of
orthogonal relay channels with multiple relays and multiple eavesdroppers (see, for example, \cite{agglalisit}).
Also, the problem of developing an additional outer bound that considers a
noiseless relay destination link remains open for the channel studied here.

\section{Acknowledgements}
The authors wish to thank Elza Erkip of Polytechnic Institute of NYU and Lifeng Lai and Ruoheng Liu of Princeton University for useful discussions related to this paper.

\appendices
\section{\label{apouterdmc}Proof of Theorem \ref{outerdmc}}

In this section, we will prove the upper bounds on the secrecy capacity for
all the three cases. Following a proof similar to that in \cite[Theorem
1]{cap_theorems:LaiElGamal}, we bound the equivocation as%

\begin{equation}
nR_{e}\leq\sum_{i=1}^{n}[I(W_{1};Y_{i}|Y^{i-1},Y_{2,i+1}^{n})-I(W_{1}%
;Y_{2,i}|Y^{i-1},Y_{2,i+1}^{n})]+n\delta_{n}. \label{A1_OB}%
\end{equation}
Now, let $J$ be a random variable uniformly distributed over $\{1,2,\cdots
,n\}$ and set $U=JY^{i-1}Y_{2,i+1}^{n}$, $V_{R}=JY_{2,i+1}W_{1}$,
$V_{D}=JY_{2,i+2}^{n}W_{1}$, $V_{2}=JY^{i-1}$, $Y_{1}=Y_{1,J}$, $Y_{2}%
=Y_{2,J}$ and $Y=Y_{J}$. We specialize the bounds in (\ref{A1_OB}) separately for
each case.

\subsection{Case 1}

From (\ref{A1_OB}), we have%

\begin{align}
R_{e}  &  \leq\frac{1}{n}\sum_{i=1}^{n}[I(W_{1};Y_{i}|Y^{i-1},Y_{2,i+1}%
^{n})-I(W_{1};Y_{2,i}|Y^{i-1},Y_{2,i+1}^{n})]+\delta_{n}\nonumber\\
&  =\frac{1}{n}\sum_{i=1}^{n}[I(W_{1},Y_{2,i+2}^{n},Y^{i-1};Y_{i}%
|Y^{i-1},Y_{2,i+1}^{n})-I(W_{1},Y_{2,i+1};Y_{2,i}|Y^{i-1},Y_{2,i+1}%
^{n})]+\delta_{n}\nonumber\\
&  =I(V_{D},V_{2};Y|U)-I(V_{R};Y_{2}|U)+\delta_{n}.%
\end{align}
Furthermore,
\begin{align}
R_{e}  &  \leq\frac{1}{n}\sum_{i=1}^{n}[I(W_{1};Y_{i}|Y^{i-1},Y_{2,i+1}%
^{n})-I(W_{1};Y_{2,i}|Y^{i-1},Y_{2,i+1}^{n})]+\delta_{n}\nonumber\\
&  =\frac{1}{n}\sum_{i=1}^{n}[I(W_{1},Y_{2,i+1}^{n},Y^{i-1};Y_{i}%
|Y^{i-1},Y_{2,i+1}^{n})-I(W_{1},Y_{2,i+1};Y_{2,i}|Y^{i-1},Y_{2,i+1}%
^{n})]+\delta_{n}\nonumber\\
&  \leq\frac{1}{n}\sum_{i=1}^{n}[I(W_{1},Y_{2,i+1}^{n},Y^{i-1};Y_{i}%
,Y_{1,i}|Y^{i-1},Y_{2,i+1}^{n})-I(W_{1},Y_{2,i+1};Y_{2,i}|Y^{i-1}%
,Y_{2,i+1}^{n})]+\delta_{n}\nonumber\\
&  =I(V_{D},V_{R},V_{2};Y,Y_{1}|V_{2},U)-I(V_{R};Y_{2}|U)+\delta_{n}.%
\end{align}
This proves the upper bound for case 1.

\subsection{Case 2}

From (\ref{A1_OB}), we have%

\begin{align}
R_{e}  &  \leq\frac{1}{n}\sum_{i=1}^{n}[I(W_{1};Y_{i}|Y^{i-1},Y_{2,i+1}%
^{n})-I(W_{1};Y_{2,i}|Y^{i-1},Y_{2,i+1}^{n})]+\delta_{n}\nonumber\\
&  =\frac{1}{n}\sum_{i=1}^{n}[I(W_{1},Y_{2,i+2}^{n},Y^{i-1};Y_{i}%
|Y^{i-1},Y_{2,i+1}^{n})-I(W_{1},Y_{2,i+2}^{n},Y^{i-1};Y_{2,i}|Y^{i-1}%
,Y_{2,i+1}^{n})]+\delta_{n}\nonumber\\
&  =I(V_{D},V_{2};Y|U)-I(V_{D},V_{2};Y_{2}|U)+\delta_{n}.%
\end{align}
Furthermore,
\begin{align}
R_{e}  &  \leq\frac{1}{n}\sum_{i=1}^{n}[I(W_{1};Y_{i}|Y^{i-1},Y_{2,i+1}%
^{n})-I(W_{1};Y_{2,i}|Y^{i-1},Y_{2,i+1}^{n})]+\delta_{n}\nonumber\\
&  =\frac{1}{n}\sum_{i=1}^{n}[I(W_{1},Y_{2,i+1}^{n},Y^{i-1};Y_{i}%
|Y^{i-1},Y_{2,i+1}^{n})-I(W_{1},Y_{2,i+2}^{n},Y^{i-1};Y_{2,i}|Y^{i-1}%
,Y_{2,i+1}^{n})]+\delta_{n}\nonumber\\
&  \leq\frac{1}{n}\sum_{i=1}^{n}[I(W_{1},Y_{2,i+1}^{n},Y^{i-1};Y_{i}%
,Y_{1,i}|Y^{i-1},Y_{2,i+1}^{n})-I(W_{1},Y_{2,i+2}^{n},Y^{i-1};Y_{2,i}%
|Y^{i-1},Y_{2,i+1}^{n})]+\delta_{n}\nonumber\\
&  =I(V_{D},V_{R},V_{2};Y,Y_{1}|V_{2},U)-I(V_{D},V_{2};Y_{2}|U)+\delta_{n}.%
\end{align}
This proves the upper bound for case 2.

\subsection{Case 3}

From (\ref{A1_OB}), we have%

\begin{align}
R_{e}  &  \leq\frac{1}{n}\sum_{i=1}^{n}[I(W_{1};Y_{i}|Y^{i-1},Y_{2,i+1}%
^{n})-I(W_{1};Y_{2,i}|Y^{i-1},Y_{2,i+1}^{n})]+\delta_{n}\nonumber\\
&  =\frac{1}{n}\sum_{i=1}^{n}[I(W_{1},Y_{2,i+2}^{n},Y^{i-1};Y_{i}%
|Y^{i-1},Y_{2,i+1}^{n})-I(W_{1},Y_{2,i+1}^{n},Y^{i-1};Y_{2,i}|Y^{i-1}%
,Y_{2,i+1}^{n})]+\delta_{n}\nonumber\\
&  =I(V_{D},V_{2};Y|U)-I(V_{R},V_{D},V_{2};Y_{2}|U)+\delta_{n}.%
\end{align}
Furthermore,
\begin{align}
R_{e}  &  \leq\frac{1}{n}\sum_{i=1}^{n}[I(W_{1};Y_{i}|Y^{i-1},Y_{2,i+1}%
^{n})-I(W_{1};Y_{2,i}|Y^{i-1},Y_{2,i+1}^{n})]+\delta_{n}\nonumber\\
&  =\frac{1}{n}\sum_{i=1}^{n}[I(W_{1},Y_{2,i+1}^{n},Y^{i-1};Y_{i}%
|Y^{i-1},Y_{2,i+1}^{n})-I(W_{1},Y_{2,i+1}^{n},Y^{i-1};Y_{2,i}|Y^{i-1}%
,Y_{2,i+1}^{n})]+\delta_{n}\nonumber\\
&  \leq\frac{1}{n}\sum_{i=1}^{n}[I(W_{1},Y_{2,i+1}^{n},Y^{i-1};Y_{i}%
,Y_{1,i}|Y^{i-1},Y_{2,i+1}^{n})-I(W_{1},Y_{2,i+1}^{n},Y^{i-1};Y_{2,i}%
|Y^{i-1},Y_{2,i+1}^{n})]+\delta_{n}\nonumber\\
&  =I(V_{D},V_{R},V_{2};Y,Y_{1}|V_{2},U)-I(V_{D},V_{2};Y_{2}|U)+\delta_{n}.%
\end{align}
This proves the upper bound for case 3.
For perfect secrecy, setting $R_1=R_e$ yields the upper bound on the secrecy capacity.
\section{\label{appdg}Proof of Theorem \ref{thpdf}: PDF for relay eavesdropper
channel}

Random Coding:

\begin{enumerate}
\item Generate $2^{n(I(X_{2};Y)-\epsilon)}$ independent and identically distributed (i.i.d.) $\mathbf{x}_{2}$'s, each with
probability $p(\mathbf{x}_{2})=\Pi_{i=1}^{n}p(x_{2i})$. Label them $x_{2}(m)$,
$m\in[1, 2^{n(I(X_{2};Y)-\epsilon)}]$.

\item For each $x_{2}(m)$, generate $2^{nR_{1}}$ i.i.d. $\mathbf{v}$'s, each with
probability $p(\mathbf{v}|x_{2}(m))=\Pi_{i=1}^{n}p(v_{i}|x_{2i}(m))$. Label
these $v(w^{\prime}|m)$, $w\in[1,2^{nR_{1}}]$.

\item For every $v(w^{\prime}|m)$, generate $2^{nR_{2}}$ i.i.d. $\mathbf{x}%
_{1}$'s, each with probability $p(\mathbf{x}_{1}|v(w^{\prime}|m))=\Pi_{i=1}%
^{n}p(\mathbf{x}_{1i}|v_{i}(w^{\prime}|m))$. Label these $x_{1}(w^{\prime
\prime}|m,w^{\prime})$, $w^{\prime\prime}\in[1,2^{nR_{2}}]$.
\end{enumerate}

\textit{Random Partition}: Randomly partition the set $\{1,2,...,2^{nR_{1}}\}$
into $2^{n(I(X_{2};Y)-\epsilon)}$ cells $S_{m}$.

\textit{Encoding}: Let $w_{i}$ be the message to be sent in block $i$ where
the total number of messages is $2^{n(R_{1}+R_{2}-I(X_{1}X_{2};Y_{2}))}$.
Further, let $g_{i}=(w_{i},l_{i})$ where $l_{i}\in\{1,2,...,2^{nI(X_{1}%
X_{2}:Y_{2})}\}$. We can further partition $g_{i}$ into two parts
$(w_{i}^{\prime},w_{i}^{\prime\prime})$ of rates $R_{1}$ and $R_{2}$
respectively. Assume that $(y_{1}(i-1),v(w_{i-1}^{\prime}|m_{i-1}%
),x_{2}(m_{i-1}))$ are jointly $\epsilon-$typical and $w_{i-1}^{\prime}\in
S_{m_{i}}$. Then the codeword $(x_{1}(w_{i}^{\prime\prime}|m_{i},w_{i}%
^{\prime}),x_{2}(m_{i}))$ will be transmitted in block $i$.

\textit{Decoding}: At the end of block $i$, we have the following:

\begin{enumerate}
\item The receiver estimates $m_{i}$ by looking at jointly $\epsilon$-typical
$x_{2}(m_{i})$ with $y_{i}$. For sufficiently large $n$, this decoding step
can be done with arbitrarily small probability of error. Let the estimate of
$m_{i}$ be $\hat{m}_{i}$.

\item The receiver calculates a set $L_{1}(y(i-1))$ of $w^{\prime}$ such that
$w^{\prime}\in L_{1}(y(i-1))$ if $(v(w^{\prime}|m_{i-1}),y(i-1))$ are jointly
$\epsilon$-typical. The receiver then declares that $w_{i-1}^{\prime}$ was
sent in block $i-1$ if $\hat{w}_{i-1}^{\prime}\in S_{m_{i}}\cap L_{1}%
(y(i-1))$. The probability that $\hat{w}_{i-1}^{\prime}=w_{i-1}^{\prime}$ with
arbitrarily high probability provided $n$ is sufficiently large and
$R_{1}<I(X_{2};Y)+I(V;Y|X_{2})-\epsilon$.

\item The receiver declares that $w_{i-1}^{\prime\prime}$ was sent in block
$i-1$ if $(x_{1}(\hat{w}_{i-1}^{\prime\prime}|\hat{m}_{i-1},\hat{w}%
_{i-1}^{\prime}),y(i-1))$ are jointly $\epsilon-$typical. $\hat{w}%
_{i-1}^{\prime\prime}=w_{i-1}^{\prime\prime}$ with high probability if $R_{2}
= I(X_{1};Y|X_{2},V)-\epsilon$ and $n$ is sufficiently large.

\item The relay upon receiving $y_{1}(i)$ declares that $\hat{w}^{\prime}$ was
received if $(v(\hat{w}^{\prime}|m_{i}),y_{1}(i),x_{2}(m_{i}))$ are jointly
$\epsilon-$typical. $w_{i}^{\prime}=\hat{w}^{\prime}$ with high probability if
$R_{1}<I(V;Y_{1}|X_{2})$ and $n$ is sufficiently large. Thus, the relay knows
that $w_{i}^{\prime}\in S_{m_{i+1}}$.
\end{enumerate}

Thus, we obtain
\begin{align}
R_{1}  & <I(X_{2};Y)+I(V;Y|X_{2})-\epsilon,\\
R_{1}  & <I(V;Y_{1}|X_{2}) \text{ \ \ and}\\
R_{2}  & =I(X_{1};Y|X_{2},V)-\epsilon.
\end{align}

Therefore, the rate of transmission from $X_{1}$ to $Y$ is bounded by%

\begin{align}
R  &  =R_{1}+R_{2}-I(X_{1}X_{2};Y_{2})\\
&  =\min\{I(X_{1};Y|X_{2},V)+I(V;Y_{1}|X_{2}),I(X_{1}X_{2}V;Y)\}-I(X_{1}%
X_{2};Y_{2}).
\end{align}

\textit{Equivocation Computation}: From \cite[Theorem 2, Equation
(41)]{cap_theorems:LaiElGamal}, we have
\begin{equation}
H(W_{1}|Y_{2})\geq H(X_{1})-I(X_{1},X_{2};Y_{2})-H(X_{1},X_{2}|W_{1},Y_{2}).
\end{equation}
Consider $H(X_{1},X_{2}|W_{1},Y_{2})$. Since we know $W_{1}$, the only uncertainty
is the knowledge of $l_{i}$ which can be decoded from $Y_{2}$ with arbitrarily
small probability of error since $l_{i}\in\{1,..,2^{nI(X_{1}X_{2};Y_{2})}\}$.
Hence,
\begin{equation}
H(W_{1}|Y_{2})\geq n(R_{1}+R_{2})-I(X_{1},X_{2};Y_{2})=nR
\end{equation}
thus giving $R_{e}=R$ and hence we get perfect secrecy.

Thus, the secrecy rate is given by
\begin{equation}
R=\min\{I(X_{1};Y|X_{2},V)+I(V;Y_{1}|X_{2}),I(X_{1}X_{2}V;Y)\}-I(X_{1}%
X_{2};Y_{2}).
\end{equation}

\section{\label{apach}Proof of Theorem \ref{thach}: PDF for relay eavesdropper
channel with orthogonal components}

From Theorem \ref{thpdf}, a secrecy rate of
\begin{equation}
R=\min\{I(X_{1};Y|X_{2},V)+I(V;Y_{1}|X_{2}),I(X_{1}X_{2}V;Y)\}-I(X_{1}%
X_{2};Y_{2})
\end{equation}
can be achieved by partial decode and forward. Let $X_{1}=(X_{R},X_{D})$ and
$V=X_{R}$ such that the input distribution is of the form $p(x_{2}%
)p(x_{R}|x_{2})p(x_{D}|x_{2})$. The achievable secrecy rate is then given by
\begin{align}
R  &  =\min\{I(X_{R}X_{D};Y|X_{2},X_{R})+I(X_{R};Y_{1}|X_{2}),I(X_{R}%
X_{D}X_{2};Y)\}-I(X_{R}X_{D}X_{2};Y_{2})\\
&  =\min\{I(X_{D};Y|X_{2},X_{R})+I(X_{R};Y_{1}|X_{2}),I(X_{D}X_{2}%
;Y)\}-I(X_{R}X_{D}X_{2};Y_{2})\\
&  =\min\{I(X_{D};Y|X_{2})+I(X_{R};Y_{1}|X_{2}),I(X_{D}X_{2};Y)\}-I(X_{R}%
X_{D}X_{2};Y_{2}) .\label{A2_B3}%
\end{align}
The equality in (\ref{A2_B3}) follows from the fact that $X_{D}-X_{2}-X_{R}$
is a Markov chain. We further specialize the bounds for the three cases based
on the receiving capability of the eavesdropper.

\subsection{Case 1}%

\begin{equation}
R=\min\{I(X_{D};Y|X_{2})+I(X_{R};Y_{1}|X_{2}),I(X_{D}X_{2};Y)\}-I(X_{R};Y_{2}).
\label{A2_B1}%
\end{equation}
The maximization of the expression to the right of the equality in
(\ref{A2_B1}) over $p(x_{D},x_{R},x_{2})=p(x_{2})p(x_{R}|x_{2})p(x_{D}|x_{2})$
is equivalent to maximizing over the more general distribution $p(x_{D}%
,x_{R},x_{2})$, and henceforth, without loss of generality we consider the
general probability distribution $p(x_{D},x_{R},x_{2})$.

We now prove that $I(X_{D};Y|X_{2})+I(X_{R};Y_{1}|X_{2})\ge I(X_{D}%
X_{R};YY_{1}|X_{2})$ which completes the proof of this part of the theorem. We have
\begin{align}
\label{rep}I(X_{D};Y|X_{2})+I(X_{R};Y_{1}|X_{2})  &  = H(Y|X_{2}%
)-H(Y|X_{2}X_{D})+I(X_{R};Y_{1}|X_{2})\nonumber\\
&  \ge H(Y|X_{2}Y_{1})-H(Y|X_{2}X_{D})+I(X_{R};Y_{1}|X_{2})\nonumber\\
&  = H(Y|X_{2}Y_{1})-H(Y|X_{2}X_{D}X_{R}Y_{1})+I(X_{R};Y_{1}|X_{2})\nonumber\\
&  = I(Y;X_{D}X_{R}|X_{2}Y_{1})+I(X_{R};Y_{1}|X_{2})\\
&  = I(Y;X_{D}X_{R}|X_{2}Y_{1})+I(X_{D};Y_{1}|X_{2}X_{R})+I(X_{R};Y_{1}%
|X_{2})\nonumber\\
&  = I(Y;X_{D}X_{R}|X_{2}Y_{1})+I(X_{R}X_{D};Y_{1}|X_{2})\nonumber\\
&  = I(YY_{1};X_{D}X_{R}|X_{2}).
\end{align}

\subsection{Case 2}%

\begin{align}
R  &  = \min\{I(X_{D};Y|X_{2})+I(X_{R};Y_{1}|X_{2}),I(X_{D}X_{2}%
;Y)\}-I(X_{D}X_{2};Y_{2}).\nonumber
\end{align}

Note that maximization of above term over $p(x_{D},x_{R},x_{2})=p(x_{2}%
)p(x_{R}|x_{2})p(x_{D}|x_{2})$ is equivalent to maximizing over general
$p(x_{D},x_{R},x_{2})$ and henceforth, without loss of generality we consider
the general probability distribution $p(x_{D},x_{R},x_{2})$.

We now prove that $I(X_{D};Y|X_{2})+I(X_{R};Y_{1}|X_{2})\ge I(X_{D}%
X_{R};YY_{1}|X_{2})$ which completes the proof of this part of the theorem. We have
\begin{align}
I(X_{D};Y|X_{2})+I(X_{R};Y_{1}|X_{2})  &  = H(Y|X_{2})-H(Y|X_{2}X_{D}%
)+I(X_{R};Y_{1}|X_{2})\nonumber\\
&  \ge H(Y|X_{2}Y_{1})-H(Y|X_{2}X_{D})+I(X_{R};Y_{1}|X_{2})\nonumber\\
&  = I(YY_{1};X_{D}X_{R}|X_{2}),
\end{align}
where the last step follows as was shown earlier in \eqref{rep}.

\subsection{Case 3}%

\begin{align}
R  &  =\min\{I(X_{D};Y|X_{2})+I(X_{R};Y_{1}|X_{2}),I(X_{D}X_{2};Y)\}-I(X_{R}%
;Y_{2}|X_{2}X_{D})-I(X_{D}X_{2};Y_{2})\nonumber\\
&  =\min\{I(X_{D};Y|X_{2})+I(X_{R};Y_{1}|X_{2}),I(X_{D}X_{2};Y)\}-I(X_{R}%
;Y_{2}|X_{2})-I(X_{D}X_{2};Y_{2}).
\end{align}

Note that maximization of above term over $p(x_{D},x_{R},x_{2})=p(x_{2}%
)p(x_{R}|x_{2})p(x_{D}|x_{2})$ is equivalent to maximizing over general
$p(x_{D},x_{R},x_{2})$.

We now prove that $I(X_{D};Y|X_{2})+I(X_{R};Y_{1}|X_{2})\ge I(X_{D}%
X_{R};YY_{1}|X_{2})$ which completes the proof of this part of the theorem. We have
\begin{align}
I(X_{D};Y|X_{2})+I(X_{R};Y_{1}|X_{2})  &  = H(Y|X_{2})-H(Y|X_{2}X_{D}%
)+I(X_{R};Y_{1}|X_{2})\nonumber\\
&  \ge H(Y|X_{2}Y_{1})-H(Y|X_{2}X_{D})+I(X_{R};Y_{1}|X_{2})\nonumber\\
&  = I(YY_{1};X_{D}X_{R}|X_{2}),
\end{align}
where the last step follows as was shown earlier in \eqref{rep}.

\end{document}